\documentclass[twocolumn,trackchanges]{aastex631}

\usepackage{CJK}
\usepackage{amsmath}
\usepackage{xcolor}
\usepackage{bm}
\usepackage{soul}
\usepackage{graphicx}

\graphicspath{figures}

\graphicspath{{./}{figures/}}

\begin{document}
\begin{CJK*}{UTF8}{gbsn}

\title{Evolution of MHD turbulence in the expanding solar wind: residual energy and intermittency}
\shortauthors{Shi et al.}

\correspondingauthor{Chen Shi}
\email{cshi1993@ucla.edu}

\author[0000-0002-2582-7085]{Chen Shi (时辰)}
\affiliation{Department of Earth, Planetary, and Space Sciences, University of California, Los Angeles \\
Los Angeles, CA 90095, USA}

\author[0000-0002-1128-9685]{Nikos Sioulas}
\affiliation{Department of Earth, Planetary, and Space Sciences, University of California, Los Angeles \\
Los Angeles, CA 90095, USA}

\author[0000-0001-9570-5975]{Zesen Huang (黄泽森)}
\affiliation{Department of Earth, Planetary, and Space Sciences, University of California, Los Angeles \\
Los Angeles, CA 90095, USA}

\author[0000-0002-2381-3106]{Marco Velli}
\affiliation{Department of Earth, Planetary, and Space Sciences, University of California, Los Angeles \\
Los Angeles, CA 90095, USA}

\author[0000-0003-2880-6084]{Anna Tenerani}
\affiliation{Department of Physics, The University of Texas at Austin, \\
     TX 78712, USA}

\author[0000-0002-2916-3837]{Victor R\'eville}
\affiliation{IRAP, Universit\'e Toulouse III - Paul Sabatier,
CNRS, CNES, Toulouse, France}



\begin{abstract}
We conduct 3D magnetohydrodynamic (MHD) simulations of decaying turbulence in the solar wind context. To account for the spherical expansion of the solar wind, we implement the expanding box model. The initial turbulence comprises uncorrelated counter-propagating Alfv\'en waves and exhibits an isotropic power spectrum.
Our findings reveal the consistent generation of negative residual energy whenever nonlinear interactions are present, independent of the normalized cross helicity $\sigma_c$ and compressibility. The spherical expansion facilitates this process. The resulting residual energy is primarily distributed in the perpendicular direction, with $[S_2(\bm{b})-S_2(\bm{u})] \propto l_\perp$ or equivalently $-E_r \propto k_\perp^{-2}$. Here $S_2(\bm{b})$ and $S_2(\bm{u})$ are second-order structure functions of magnetic field and velocity respectively.
In most runs, $S_2(\bm{b})$ develops a scaling relation $S_2(\bm{b}) \propto l_\perp^{1/2}$ ($E_b \propto k_\perp^{-3/2}$). In contrast, $S_2(\bm{u})$ is consistently shallower than $S_2(\bm{b})$, which aligns with in-situ observations of the solar wind.
We observe that the higher-order statistics of the turbulence, which act as a proxy for intermittency, depend on the initial $\sigma_c$ and are strongly affected by the expansion effect. Generally, the intermittency is more pronounced when the expansion effect is present.
Finally, we find that in our simulations although the negative residual energy and intermittency grow simultaneously as the turbulence evolves, the causal relation between them seems to be weak, possibly because they are generated on different scales.

\end{abstract}



\section{Introduction} \label{sec:intro}
It has long been observed that solar wind is a highly turbulent plasma system with fluctuations on a wide range of scales \citep[see the review by][and references therein]{bruno2013solar}. 
Studying the solar wind turbulence is of great importance because turbulence is an important power source for the heating and acceleration of solar wind \citep{cranmer2007self,verdini2009turbulence,lionello2014validating,cranmer2015role,van2016heating,shoda2019three,reville2020role,magyar2021three,halekas2023quantifying,rivera2024situ}.

In the last decades, significant progresses have been made on observations, numerical simulations, and theories of the solar wind turbulence. 
Satellite observations reveal that in fast solar wind, the turbulence is usually highly Alfv\'enic, dominated by outward propagating Alfv\'en waves \citep{belcher1971large}, while in slow solar wind, Alfv\'enicity of the turbulence is typically lower than the fast wind, but can be quite high in certain intervals \citep{d2015origin,d2019slow}, 
especially in the nascent solar wind as observed by Parker Solar Probe \citep{panasenco2020exploring,parashar2020measures}. 

Since the compressible fluctuation is typically small in the solar wind, with $\delta n / n \lesssim 0.2$ \citep{shi2021alfvenic} where $n$ is the average plasma density and $\delta n$ is the fluctuation amplitude of the density, the solar wind turbulence is treated as an incompressible MHD system in most theoretical and modeling works, where two Els\"asser variables $\bm{z^\pm} = \bm{u} \mp \bm{b}$, which are linear combinations of the velocity $\bm{u}$ and magnetic field $\bm{b}$ (in Alfv\'en speed) and represent the two counter-propagating Alfv\'en wave populations, are analyzed. 
A number of phenomenological models have been developed for the incompressible MHD turbulence. The weak ($|\delta \bm{b}| / B \ll 1$), isotropic (in $\bm{k}$ space), balanced ($z^+ \sim z^-$) turbulence model \citep{iroshnikov1964turbulence,kraichnan1965inertial} predicts a 1D power spectrum $E_{1D} \propto k^{-3/2}$. 
The weak, anisotropic, balanced model \citep{goldreich1997magnetohydrodynamic} predicts a 1D power spectrum $E_{1D} \propto k_\perp^{-2}$. 
For strong, anisotropic turbulence, ``critical balance'' theory \citep{goldreich1995toward}, which balances the linear propagation timescale and the nonlinear eddy turnover timescale, predicts $E_{1D} \propto k_\perp^{-5/3}$. 
Based on the critical balance theory, scale-dependent dynamic alignment model \citep{boldyrev2005spectrum,perez2007weak} allows the spectral slope to be variable depending on how much the two Els\"asser variables are aligned with each other. 
Compared with the balanced turbulence, imbalanced turbulence is more difficult to describe, and no simple phenomenological model has been established so far \citep{dobrowolny1980properties,dobrowolny1980fully,grappin1983dependence,lithwick2003imbalanced,lithwick2007imbalanced,beresnyak2010scaling}. 
Incompressible simulations conducted by \citet{perez2009role} show that in strong turbulence the two imbalanced Els\"asser variables may have similar power spectra despite of different amplitudes, while simulations conducted by \citet{beresnyak2009structure} show that the two Els\"asser variables have very different structures.

As the phenomenological models and previous numerical simulations have successfully explained some of the satellite observations, many mysteries still remain. One of the most outstanding problems is the prevailing negative residual energy, i.e. an excess of magnetic energy over the kinetic energy, in the solar wind turbulence \citep{chen2013residual,chen2020evolution,shi2021alfvenic,sioulas2023magnetic}. 
Many theoretical works have been carried out \citep{muller2005spectral,yokoi2007application,wang2011residual,boldyrev2012residual,gogoberidze2012generation,howes2013alfven,dorfman2024residual} to explain the generation of negative residual energy but they are not fully self-consistent and do not give consistent results, e.g. on spectral slope of the residual energy.
In addition, although most of the phenomenological models assume self-similarity, intermittency plays a non-negligible role in MHD turbulence as it undermines the self-similarity assumption \citep{chandran2015intermittency,mallet2017statistical,wu2023intermittency}. 
Besides, intermittency is an important way of energy dissipation and is observed to be directly correlated with plasma heating in the solar wind \citep{sioulas2022preferential,sioulas2022statistical,phillips2023association}.


\begin{figure*}[htb!]
    \centering
    \includegraphics[width=\linewidth]{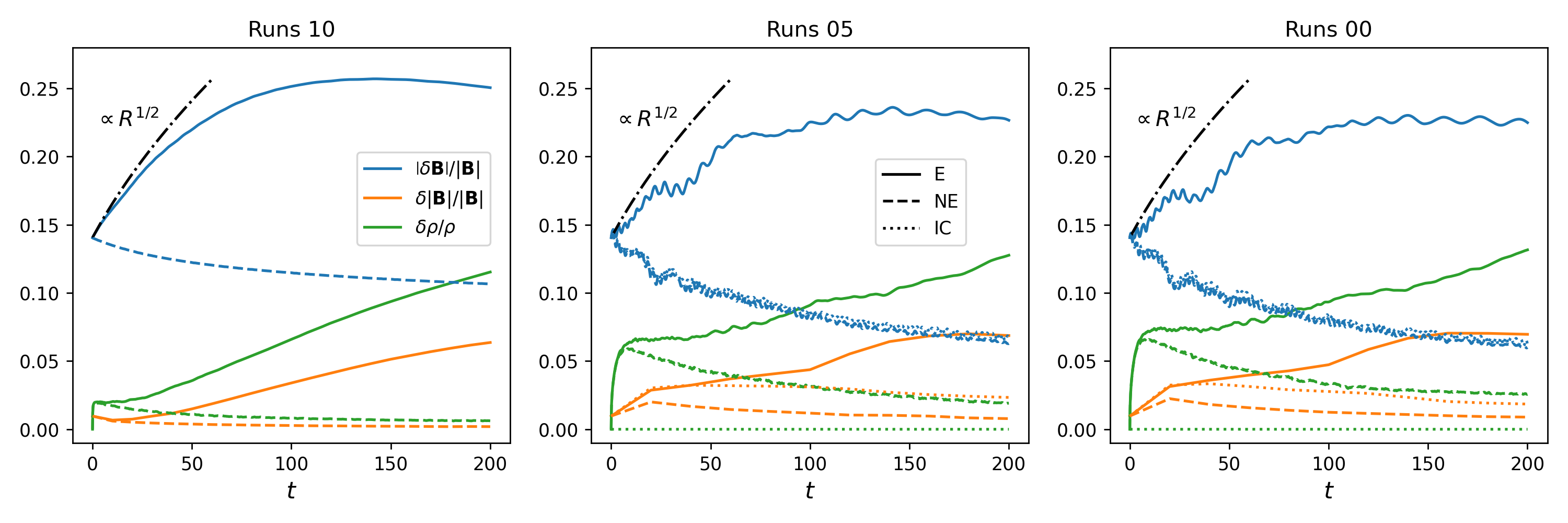}
    \caption{Time evolution of $ \left|\delta \bm{B} \right|/ \left|\bm{B} \right|$ (blue), $\delta \left| \bm{B} \right|/\bm{B}$ (orange), and $\delta \rho / \rho$ (green) in different runs. Here $\delta \bm{B}$ is the root-mean-square (RMS) of the magnetic field vector, $\delta \left| \bm{B} \right|$ is the RMS of the magnetic field magnitude, $\delta \rho$ is the RMS of the density. $\left| \bm{B} \right|$ is the amplitude of the average magnetic field, and $\rho$ is the average density. Left, middle, and right panels show Runs 10, Runs 05, and Runs 00 respectively. The black dashed-dotted lines show the growth of $\left| \delta \bm{B} \right|/ \left| \bm{B} \right|$ predicted by WKB theory, i.e. $\propto R^{1/2}$.}
    \label{fig:rms_density_modB_Bvec}
\end{figure*}

In this study, we investigate, through 3D MHD simulations, turbulence evolution in the solar wind context with a focus on the residual energy and intermittency.
Expanding-box-model (EBM) \citep{grappin1996waves,dong2014evolution,tenerani2017evolving,shi2020propagation,shi2022influence,grappin2022modeling} was implemented because the spherical expansion of the solar wind may significantly change the turbulence evolution as it leads to anisotropic decay of different components of the magnetic field and velocity and may result in mode conversion between different wave modes \citep{huang2022conservation}. 
The paper is organized as follows. In Section \ref{sec:model}, we describe the simulation setup. In Section \ref{sec:result}, we present the simulation results. In Section \ref{sec:discussion} we discuss the relation between the residual energy and intermittency. In Section \ref{sec:summary}, we summarize this work.

\section{Simulation setup}\label{sec:model}
We use the \texttt{LAPS} code\footnote{https://github.com/chenshihelio/LAPS}, which is a 3D pseudo-spectral compressible MHD code with EBM, to conduct the simulations. 
The algorithm of the code is described in detail in \citep{shi2024laps}. In all the simulations, the domain is a rectangular box with initial size $(5R_s)^3$ ($R_s$ is the solar radius) and grid number $512^3$. Besides de-aliasing in $\bm{k}$-space, explicit resistivity $\eta=2\times 10^{-5}$ and viscosity $\nu=2\times 10^{-5}$ are implemented to maintain numerical stability. We note that, because the code is based on MHD equation in conservation form, the viscosity is implemented as $\partial_t(\rho \bm{u})_{\bm{k}} \sim -k^2 \nu (\rho \bm{u})_{\bm{k}}$ where $(\rho \bm{u})_{\bm{k}}$ is Fourier mode $\bm{k}$ of the conserved variable $\rho \bm{u}$.

The initial fields consist of uniform background and fluctuations. The background fields are $\rho_0 = B = 1$, $P_0 = 0.1006$, with normalization units $\bar{n} = 200$ cm$^{-3}$, $\bar{B} = 250$ nT, and subsequently $\bar{P}= \bar{B}^2/\mu_0 = 49.7$ nPa where $\mu_0$ is the permeability. The background magnetic field is within the equatorial plane ($x-y$ plane) and has an angle of $8.1^\circ$ with respect to the radial direction ($\hat{e}_x$), so that in simulations with expansion this angle increases to about $45^\circ$ at 1AU. The adiabatic index is $\gamma=1.5$ instead of $5/3$ to prevent the plasma temperature from cooling too fast in the runs with expansion. This choice only slightly modifies the thermodynamics and is not expected to impact our results significantly. 

Fluctuations of velocity and magnetic field with 3D isotropic power spectra are added on top of the background fields. These initial fluctuations are added on largest scales contained in the simulation domain such that $|k| \in [1/L,32/L]$ where $L=5R_s$ is the domain size. The reduced 1D spectra of the initial fluctuations roughly follow $|k|^{-1.3}$.
Consequently, we are not able to observe the shallow ``$1/f$ range'' which is usually observed in the solar wind \citep[e.g.][]{matteini20181,huang2023new} and may be generated due to the inverse cascade \citep{chandran2018parametric,meyrand2023reflection}. 
We note that there are no forcing terms in the model equations, i.e. the turbulence is decaying. 
One should be aware that, with the expansion effect, in addition to dissipation, the turbulence also decays due to energy exchange with the background solar wind.
The initial fluctuations are Alfv\'enic: for any wave mode $\bm{k}$, there is $\bm{b_k} \propto (\bm{k} \times \bm{B})$, where $\bm{b_k}$ is the magnetic field fluctuation of wave-vector $\bm{k}$. 
Usually, we use the normalized cross helicity $\sigma_c$ and normalized residual energy $\sigma_r$ to measure the Alfv\'enicity of the turbulence, and they are defined as 
\begin{equation}
    \sigma_c = \frac{E_+ - E_-}{E_+ + E_-}, \sigma_r = \frac{E_k - E_b}{E_k + E_b}
\end{equation}
where $E_{\pm}$ represent the energy of the outward/inward Alfv\'en waves ($\bm{z^\pm}$) and $E_{k,b}$ represent the kinetic and magnetic energies of the fluctuations. 
We note that $\sigma_c$ measures the correlation between the velocity and magnetic field, and $\sigma_r$ measures the correlation between the two Els\"asser variables.
At initialization, we control $\sigma_c$ by varying the correlation between the velocity fluctuation and magnetic field fluctuation, and we keep $\sigma_r$ to be exactly zero. 

\begin{figure*}[htb!]
    \centering
    \includegraphics[width=\hsize]{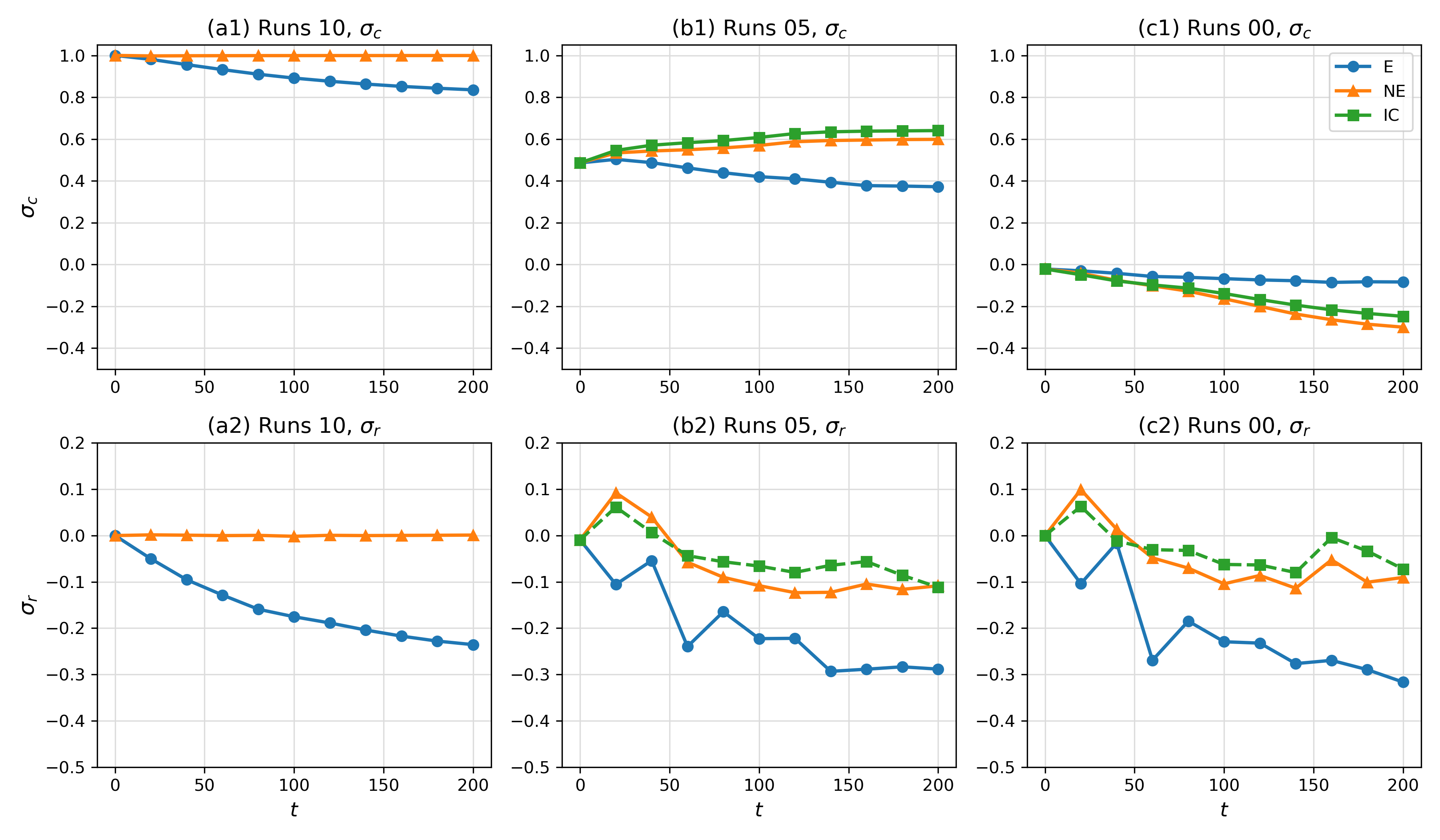}
    \caption{Top panels: Evolution of $\sigma_c$ in Runs 10 (a1), Runs 05 (b1), and Runs 00 (c1). Blue circles are runs with expansion, orange triangles are runs without expansion, and green squares are incompressible-MHD runs without expansion. Bottom panels have the same format with top panels but show the evolution of $\sigma_r$.}
    \label{fig:evolution_sigma_c_sigma_r}
\end{figure*}

The root-mean-square (RMS) of the magnetic field fluctuation is $b_{rms}/B \approx 0.14$ for all the runs. Thus, the nonlinear eddy turnover time is estimated to be $\tau_{nl} \sim L/2\pi b_{rms} \approx 5.7$ and the effective Reynolds number $Re$ (and Lundquist number $S$) is $Re = S  \approx L b_{rms} / \nu \approx 3.5 \times 10^4$. We note that because the background plasma has quite low $\beta$($\approx 0.2$) as we want the configuration to be close to the realistic solar wind in the inner heliosphere \citep{artemyev2022ion}, the fluctuation level cannot be too strong, otherwise the simulation will be unstable due to formation of shocks.  
In contrast, \citep{dong2014evolution} and \citep{grappin2022modeling} added strong turbulence (with $b_{rms}/B \approx 1$) to their simulations by adopting large $\beta$. 
The initial turbulence Mach number in our simulations is $M_s = u_{rms} / C_s \approx 0.36$, where $u_{rms} = b_{rms} = 0.14$ and $C_s = \sqrt{\gamma P_0 / \rho_0} = 0.39$ is the sound speed.

\begin{figure*}[htb!]
    \centering
    \includegraphics[width=\hsize]{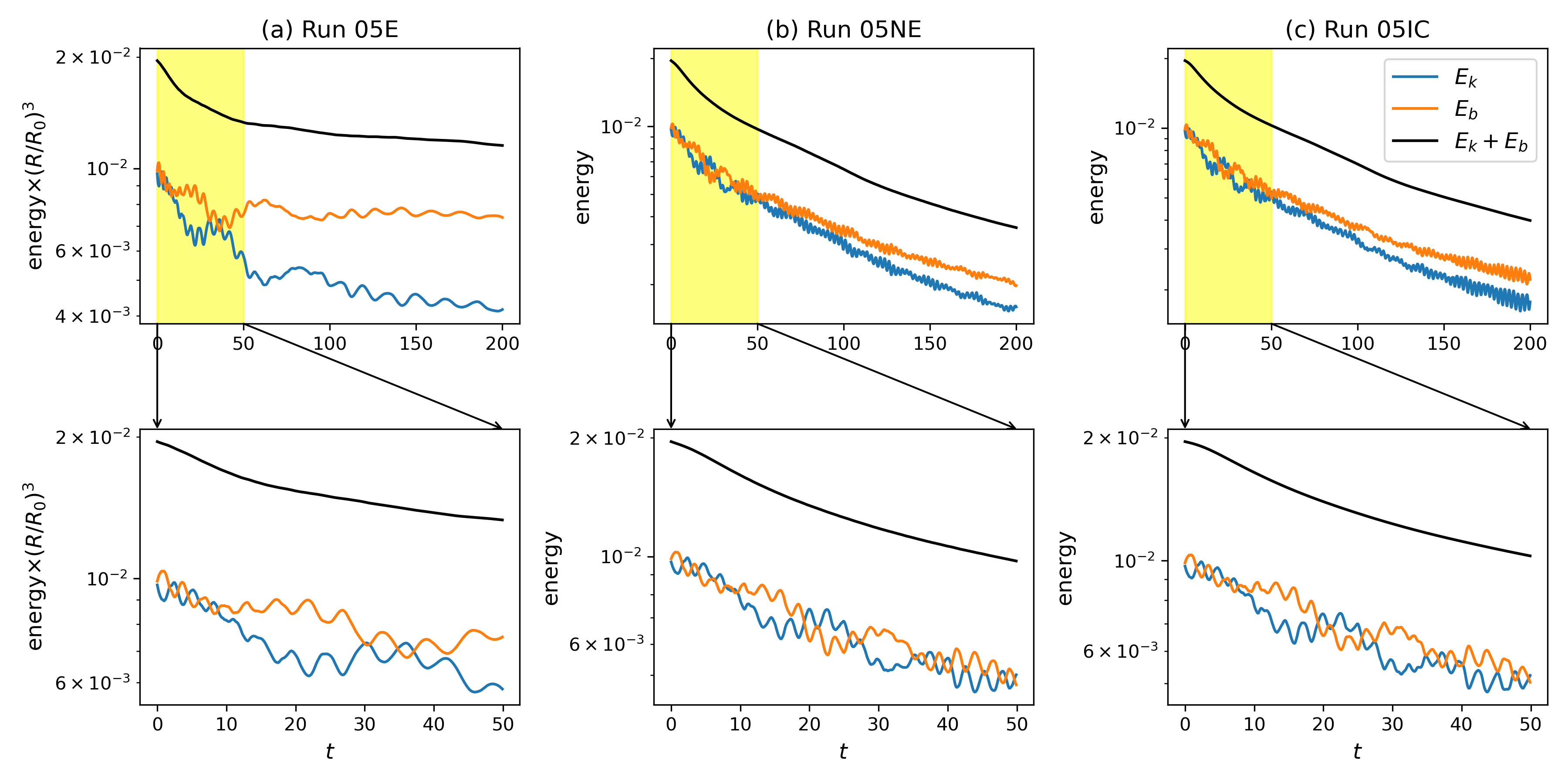}
    \caption{Evolution of kinetic energy (blue), magnetic energy (orange), and total energy (black) with high-resolution output for Run 05E (a), Run 05NE (b), and Run 05IC (c). Bottom row shows blow-ups of the yellow-shaded regions in the top row. For Run 05E, we have multiplied the energies by $(R/R_0)^3$ to compensate the energy decay due to expansion based on WKB theory.}
    \label{fig:evolution_EkEb}
\end{figure*}

In the runs with expansion, the initial radial location of the simulation domain is $R_0 = 30R_s$, and the radial speed of the box is $U_r=1.167$ with normalization unit $\bar{U} = \bar{B}/\sqrt{\mu_0 m_p \bar{n}} = 385.6$ km/s where $m_p$ is the proton mass. 
We carry out six compressible-MHD runs, which are divided into three groups: Runs (10E, 10NE), Runs (05E, 05NE), and Runs (00E, 00NE). Here ``E'' and ``NE'' stand for ``expansion'' and ``no expansion'' respectively. Runs 10 have $\sigma_{c,0} = 1$, Runs 05 have $\sigma_{c,0} \approx 0.5$, and Runs 00 have $\sigma_{c,0} \approx 0$ where $\sigma_{c,0}$ is the initial normalized cross helicity.
In addition, based on Runs 05NE and Run 00NE, we carry out two extra runs using the incompressible version of the \texttt{LAPS} code. These two runs are labeled as Run 05IC and Run 00IC.

\begin{figure*}[htb!]
    \centering
    \includegraphics[width=0.92\hsize]{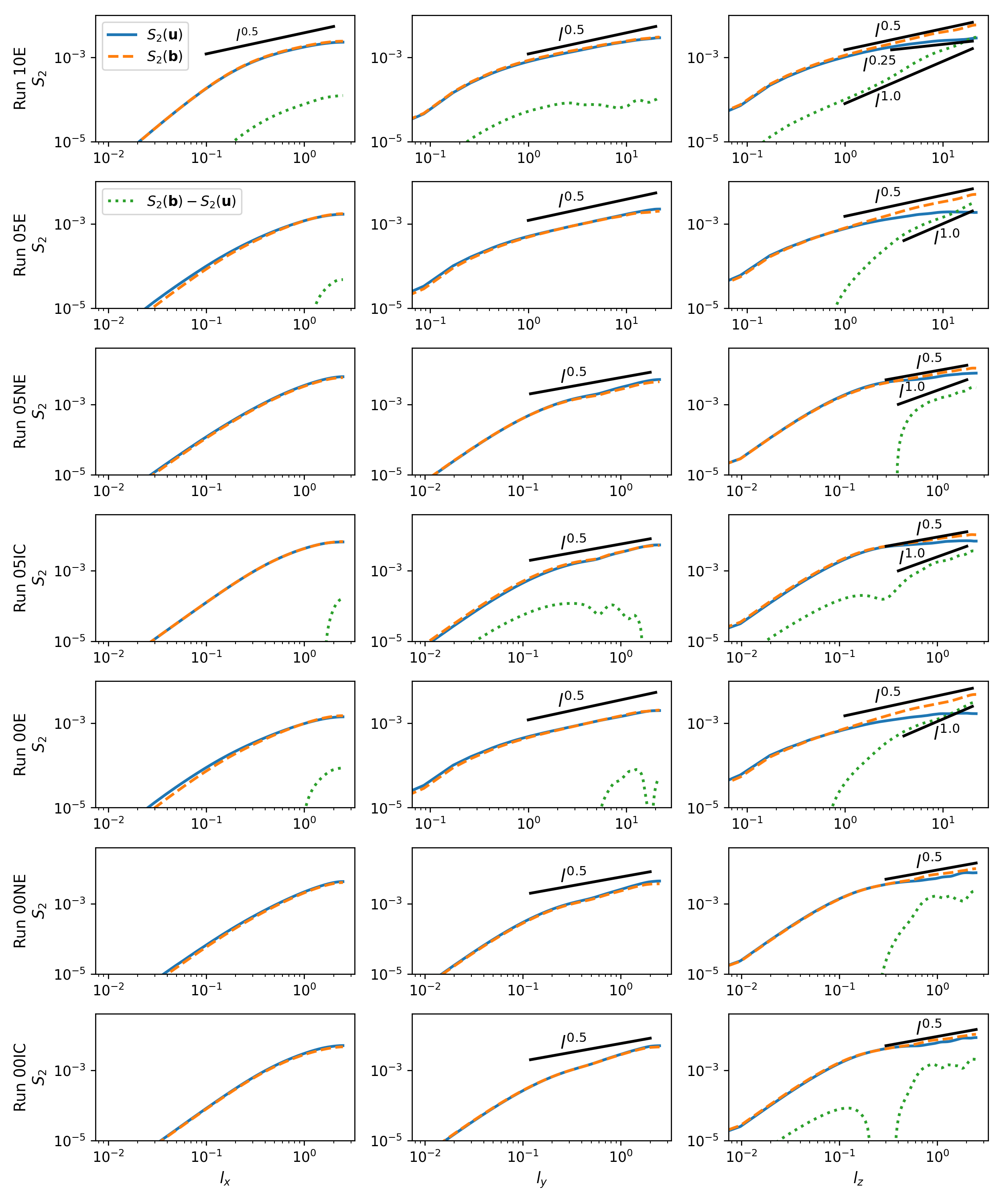}
    \caption{Second-order structure functions of velocity $S_2(\bm{u})$ (blue), magnetic field $S_2(\bm{b})$ (orange dashed), and the difference between them $S_2(\bm{b}) - S_2(\bm{u})$ (green dotted) at the end of simulations ($t=200$). Left to right columns are $\bm{l} = l \hat{e}_x$, $\bm{l} = l \hat{e}_y$, and $\bm{l} = l \hat{e}_z$ respectively. Top to bottom rows are Runs 10E, 05E, 05NE, 05IC, 00E, 00NE, and 00IC respectively.}
    \label{fig:S2_ub}
\end{figure*}

\section{Result}\label{sec:result}

\subsection{Compressibility}\label{sec:compressibility}

Figure \ref{fig:rms_density_modB_Bvec} shows the time evolution of the normalized fluctuation levels of the magnetic field vector (blue), magnetic field magnitude (orange), and density (green) in different runs. 
Here the fluctuation level is defined as the RMS of a specific quantity. We use the magnitude of the average magnetic field and the average density for normalization.
In each panel, the solid curves correspond to the run with expansion, the dashed curves correspond to the run without expansion, and the dotted curves correspond to the incompressible run.

Comparing the three panels, we find that the initial $\sigma_c$ does not have a strong impact on the evolution of the three parameters.
In runs with expansion, $ \left| \delta \bm{B}  \right| / \left| \bm{B} \right|$ increases at the beginning and then slowly decays after saturation. 
The increase is attributed to the slower decay of the Alfv\'en wave amplitude than the background magnetic field in the expanding solar wind \citep{belcher1971alfvenic,hollweg1974transverse}.
However, we note that the increase is slower than the WKB prediction, i.e. $\propto R^{1/2}$, as shown by the black dashed-dotted curves in Figure \ref{fig:rms_density_modB_Bvec}, because of the nonlinear energy cascade.
The latter decrease is because of the dissipation of the turbulence energy, as can be seen in the runs without expansion.
Because of the pressure-imbalance at the initial status, compressible fluctuations are generated soon after the simulations start.
Interestingly, the density fluctuation is stronger in runs with lower $\sigma_c$, indicating that the nonlinear interaction plays an important role in the generation of compressible fluctuations.
Similar to $ \left| \delta \bm{B}  \right| / \left| \bm{B} \right|$, the normalized density fluctuation increases with time in runs with expansion but decreases in runs without expansion.
The magnetic compressibility $ \delta \left|  \bm{B}  \right| / \left| \bm{B} \right|$ follows a similar trend with $\delta \rho / \rho$, i.e. it increases with time in runs with expansion and decreases in runs without expansion.
We note that, in a recent study by \citet{matteini2024alfvenic}, it was found that the magnetic field magnitude evolves towards uniform in 2D EBM hybrid simulations of balanced turbulence. 
This is, however, not observed in our MHD simulations, which show that the magnetic compressibility continues to increase with time.
This discrepancy may imply that kinetic physics is necessary to produce the spherically polarized Alfv\'en waves that dominate the solar wind turbulence.
In the EBM-MHD simulations conducted by \citet{squire2020situ}, the magnetic compressibility is observed to be small as the fluctuation level of the magnetic field vector becomes similar to the background magnetic field, accompanied by the generation of rotational discontinuities \citep{vasquez1998formation}. 
However, our simulations are not directly comparable to those by \citet{squire2020situ} because the amplitude of the fluctuations is relatively small. 
A detailed analysis of the nature of the compressible fluctuations in the EBM-MHD simulations is necessary but will be left for a future study.

\subsection{Evolution of $\sigma_c$ and $\sigma_r$}\label{sec:evolution_sigma_c_sigma_r}

In Figure \ref{fig:evolution_sigma_c_sigma_r}, we show the time evolution of $\sigma_c$ (top) and $\sigma_r$ (bottom) in Runs 10 (a), Runs 05 (b), and Runs 00 (c) respectively. Blue lines with circles are runs with expansion, orange lines with triangles are runs without expansion, and green lines with squares are incompressible runs without expansion. 
With expansion, $|\sigma_c|$ gradually decreases in imbalanced turbulence (Runs 10E \& 05E), because of the reflection of the outward propagating Alfv\'en waves due to the inhomogeneity of the background fields \citep{heinemann1980non,velli1991waves}. 
Without expansion, Run 10NE does not evolve because nonlinear interaction is absent in the exactly Alfv\'enic status ($\sigma_{c,0} = 1$) . 
In Run 05NE and Run 05IC, $\sigma_c$ increases with time, possibly because of the ``dynamic alignment'' \citep{dobrowolny1980properties,dobrowolny1980fully}, i.e. an initially imbalanced turbulence tends to evolve toward purely Alfv\'enic status because the energy decay rates of the two counter-propagating Alfv\'en wave populations are similar. 
For the (nearly)-balanced turbulence (Runs 00), whether $\sigma_c$ evolves toward positive or negative is very sensitive to the initial condition. In the simulations conducted here, the initial $\sigma_c$ is slightly negative.
Hence, in Run 00NE and Run 00IC, $\sigma_c$ decreases to more and more negative values. 
In Run 00E, $\sigma_c$ remains negative but stays at very low absolute values. This is because of the competition between the dynamic alignment which tends to increase $|\sigma_c|$ and the expansion effect which tends to decrease $|\sigma_c|$. 
By comparing the green and orange lines, we can see that the evolution of $\sigma_c$ in the incompressible runs (Runs 05IC and 00IC) does not show big differences from that in the compressible runs (Runs 05NE and Runs 00NE).

\begin{figure*}[htb!]
    \centering
    \includegraphics[width=0.92\hsize]{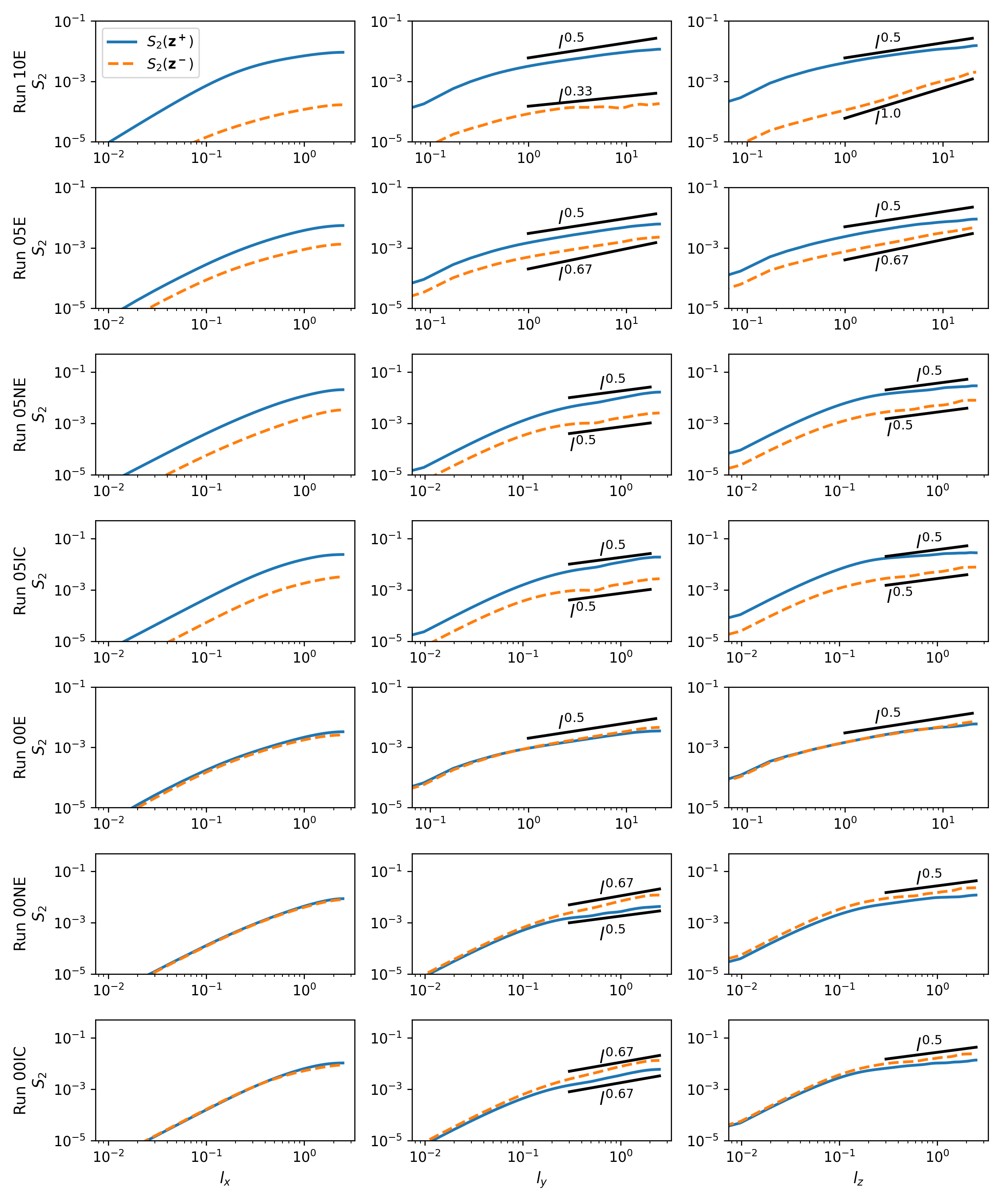}
    \caption{Second-order structure functions of Els\"asser variables $S_2(\bm{z^+})$ (blue) and $S_2(\bm{z^-})$ (orange dashed) at the end of simulations ($t=200$). Left to right columns are $\bm{l} = l \hat{e}_x$, $\bm{l} = l \hat{e}_y$, and $\bm{l} = l \hat{e}_z$ respectively. Top to bottom rows are Runs 10E, 05E, 05NE, 05IC, 00E, 00NE, and 00IC respectively.}
    \label{fig:S2_zpm}
\end{figure*}

$\sigma_r$ evolves toward negative values in all the runs except for Run 10NE, indicating that nonlinear interaction naturally generates negative residual energy \citep{grappin1983dependence,yokoi2007application,gogoberidze2012generation,wang2011residual,boldyrev2012residual,howes2013alfven,dorfman2024residual}, consistent with the prevailing negative residual energy observed in the solar wind \citep{chen2013residual,chen2020evolution,shi2021alfvenic}. 
In addition, it is clear that $\sigma_r$ decays faster and to more negative values in runs with expansion than those without expansion. This may be attributed to the fact that the expansion-induced decay of magnetic energy is slower than that of kinetic energy \citep{dong2014evolution,shi2022influence}, i.e. 
\begin{equation}
    b_\perp^2 \propto R^{-2},  \rho u_\perp^2 \propto R^{-4}
\end{equation}
in the absence of any coupling between $\bm{b}$ and $\bm{u}$. This effect is significant for non-propagating, perpendicular modes with $\bm{k} \perp \bm{B_0}$ \citep{meyrand2023reflection}.
Similar to $\sigma_c$, $\sigma_r$ in the incompressible runs does not differ much from the compressible runs.

As an example, in Figure \ref{fig:evolution_EkEb}, we show the time evolution of kinetic (blue) and magnetic (orange) energies in Run 05E, Run 05NE, and Run 05IC respectively with high time resolution. Total energy is also shown as the black curves. For Run 05E we have multiplied the energies by $(R/R_0)^3$ to compensate the energy decay due to solar wind expansion \citep[the WKB theory,][]{belcher1971alfvenic}.
We observe high-frequency oscillations of $E_k$ and $E_b$, which are anti-correlated so that the total energy does not oscillate. 
This oscillation is clearly a result of the wave propagation effect \citep{wang2011residual}. From panel (b) or (c), we can estimate the oscillation period is roughly $T \approx 2.5 \approx L_x / 2B_0$, and from panel (a) we see that the period increases gradually because the expansion increases the crossing time of Alfv\'en waves through the simulation domain. Figure \ref{fig:evolution_EkEb} indicates that the magnetic energy excess is built up over multiple wave crossing times.


\subsection{Second-order structure functions}
We then investigate the second-order structure functions of different fields. The $q$-th order structure function of a field $\bm{b}(\bm{x})$ is defined as
\begin{equation}
    S_q(\bm{b},\bm{l}) = \left< \left|\bm{b}(\bm{x}+\bm{l}) -\bm{b}(\bm{x}) \right|^q   \right>_{\bm{x}}
\end{equation}
where $\bm{l}$ is a given spatial increment, and $\left< \right>_{\bm{x}}$ means ensemble average or equivalently average over the whole simulation domain. 
The second-order structure function $S_2$ measures the mean-square value of the fluctuation on scale $\bm{l}$. One important relation is that, if $S_2$ scales exponentially with the spatial increment such that $S_2 \propto l^{\alpha}$, the power spectrum of the field obeys the scaling $E \propto k^{-(\alpha + 1)}$ \citep{montroll19821}.

\begin{figure*}[htb!]
    \centering
    \includegraphics[width=\hsize]{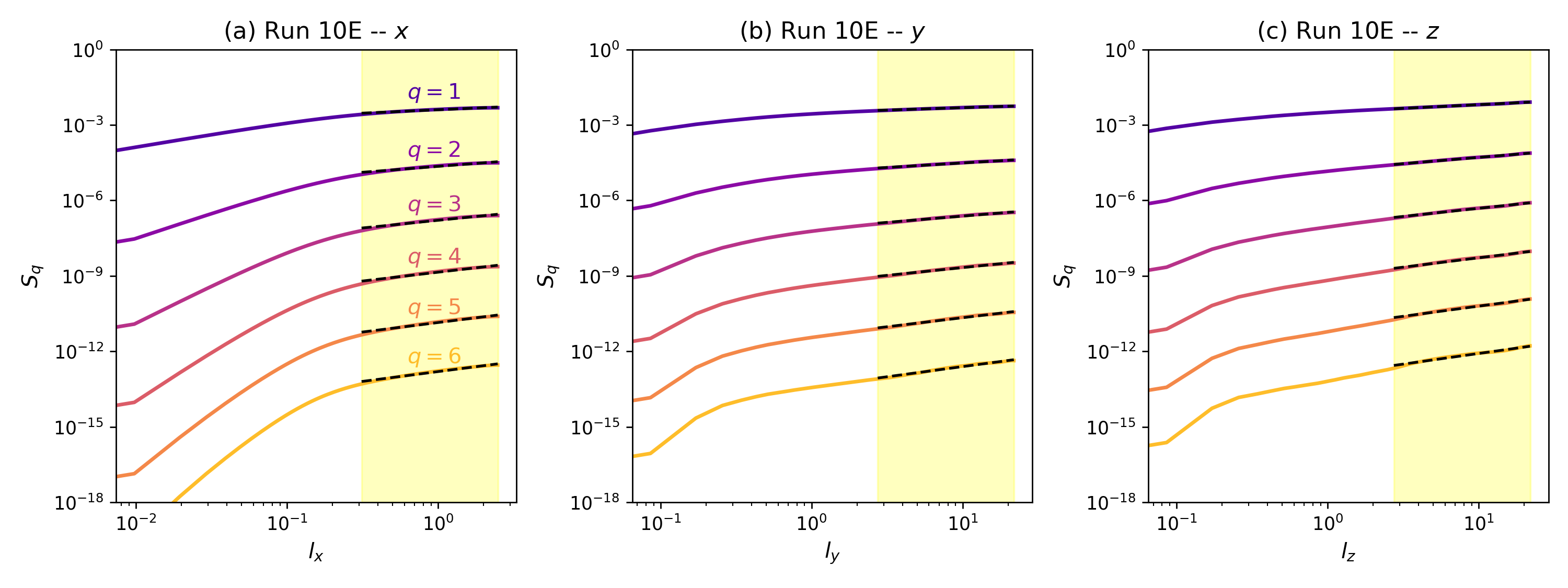}
    \caption{Panels (a)-(c): Structure functions of magnetic field $S_q(\bm{b},\bm{l})$ for $\bm{l} = l_x \hat{e}_x$, $\bm{l} = l_y \hat{e}_y$, and $\bm{l} = l_z \hat{e}_z$ at the end of Run 10E ($t=200$). Curves for $q=1-6$ are plotted. Yellow shades mark the scale ranges used for fitting the structure functions. The linear-fitting result is shown by the black dashed lines.}
    \label{fig:Sq_mag_Run10E}
\end{figure*}

Figure \ref{fig:S2_ub} shows $S_2$ of velocity (blue), magnetic field (orange dashed) and negative residual energy (magnetic energy minus kinetic energy, green dotted) at the end of the simulations ($t=200$), when turbulence has evolved a sufficient time.
The left, middle, and right columns show structure functions with $\bm{l}$ along $x$, $y$, and $z$. 
From top to bottom rows are Runs 10E, 05E, 05NE, 05IC, 00E, 00NE, and 00IC respectively.
Run 10NE is not shown because of the absence of nonlinear evolution.
Anisotropy among the three directions is clearly observed in all the runs. 
Along $l_x$ (radial and initially quasi-parallel), no clear power-law relation is established. Along $l_y$, an extended power-law part with a slope slightly smaller than 0.5 forms in all the runs for both $\bm{u}$ and $\bm{b}$. 
Along $l_z$, we get $S_2(\bm{b}) \propto l_z^{0.5}$ and $S_2(\bm{u})$ shallower than $S_2(\bm{b})$. This is consistent with satellite observations \citep{chen2020evolution,shi2021alfvenic}, though the spectral slopes from the simulations are systematically shallower than satellite observations which show that the magnetic field and velocity power spectra have slopes of $-5/3$ and $-3/2$ on average \citep{chen2013residual}. 
By comparing different rows, one can see that $S_2(\bm{b}) \propto l_z^{0.5}$ holds for all the runs, independent of the initial $\sigma_c$ and the expansion effect, while  $S_2(\bm{u})$ is clearly shallower in the runs with expansion than runs without expansion.
Residual energy is generated mainly along the $z$ direction, i.e. the direction perpendicular to both the background magnetic field and radial direction. 
The negative residual energy has a power-law scaling $S_2 \propto l_z$, i.e. $-E_r \propto k_z^{-2}$ in Runs 10E, 05E, 05NE, 05IC, and 00E. This spectral slope is consistent with the WIND observation \citep{chen2013residual} as well as the prediction given by the Eddy Damped Quasi Normal Markovian (EDQNM) model of isotropic MHD turbulence \citep{grappin1983dependence}\footnote{In \citep{grappin1983dependence}, the sign of the residual energy is not defined.}, which is verified by 3D MHD simulations with zero mean magnetic field \citep{grappin2016alfven}, and is consistent with the model by \citet{boldyrev2012residual} for anisotropic strong balanced turbulence. 
However, we note that there is so far no self-consistent theory for the spectral slope of residual energy, and different models can give different results. 
For example, \citet{wang2011residual} shows, through analytic calculation, that in the weak turbulence scenario negative residual energy is produced and follows $-E_r \propto k_\perp^{-1}$. 
The EDQNM model for strong, anisotropic MHD turbulence \citet{gogoberidze2012generation} also shows that negative residual energy arises due to nonlinear interaction but its spectrum follows $-E_r \propto k_\perp^{-5/3}$.
Our results reveal that the generation of residual energy is strongly anisotropic with the presence of a finite background magnetic field. 
Even with a small mean field, e.g. along the $y$-direction in Run 05NE, residual energy is barely observed, possibly due to the ``Alfv\'en effect'' that dissipates the residual energy \citep{kraichnan1965inertial,grappin2016alfven}.

\begin{figure*}[htb!]
    \centering
    \includegraphics[width=\hsize]{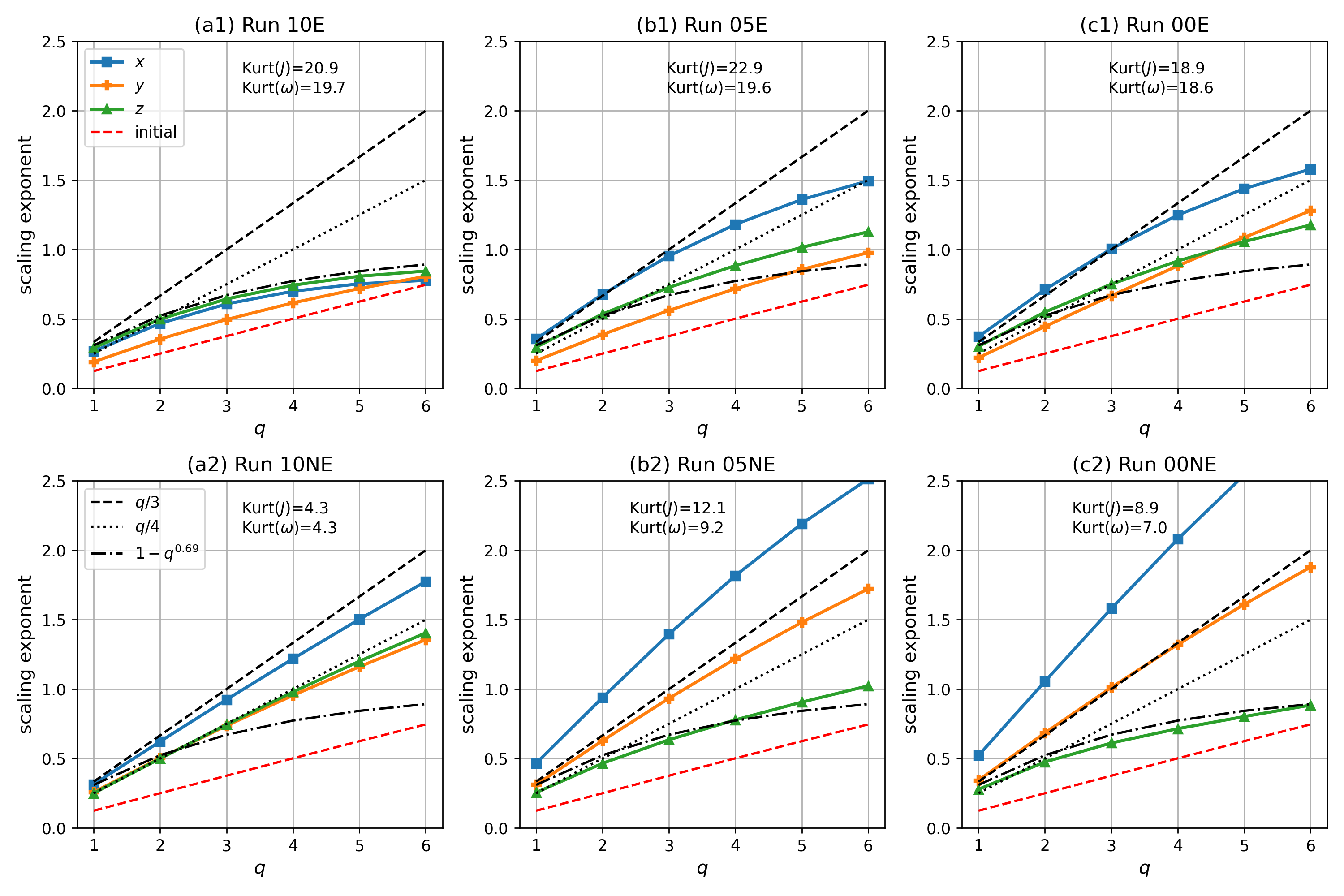}
    \caption{Scaling exponents as functions of $q$ along $x$ (blue square), $y$ (orange cross), and $z$ (green triangle) directions at the end of different runs. In each panel, red dashed line shows the initial status ($t=0$). The three black lines are $q/3$, $q/4$, and $1-q^{0.69}$ for reference.}
    \label{fig:scaling_exponent}
\end{figure*}

In Figure \ref{fig:S2_zpm}, we show $S_2(\bm{l})$ of $\bm{z^+}$ (blue) and $\bm{z^-}$ (orange dashed) at the end of different runs. 
Similar to Figure \ref{fig:S2_ub}, for $\bm{l} = l \hat{e}_x$ (quasi-parallel to $\bm{B_0}$), the structure function does not evolve much. For $\bm{l} = l \hat{e}_y$, the structure function is determined by a mix of parallel and perpendicular effects. Therefore, we will focus on the right column, i.e. $\bm{l} = l \hat{e}_z$.
$S_2(\bm{z^+})$ has very similar shapes, i.e. with a slope slightly shallower than $0.5$, in all the runs. 
The slope of $S_2(\bm{z^-})$, however, behaves very differently from $S_2(\bm{z^+})$. 
For Runs 05NE, 05IC, 00NE, and 00IC, the slope is roughly $0.5$, but for runs with expansion, it is strongly affected by $\sigma_c$. 
For Run 10E, $S_2(\bm{z^-}) \propto l_z$, for Run 05E, $S_2(\bm{z^-}) \propto l_z^{2/3}$, and for Run 00E, $S_2(\bm{z^-}) \propto l_z^{0.5}$. That is to say, the perpendicular spectrum of $\bm{z^-}$ is steeper as the turbulence gets more imbalanced. 
In previous numerical works without expansion effect \citep{perez2009role,perez2012energy}, the $\bm{z^-}$ spectral slope is $E \sim k_\perp^{-1.5}$ for both balanced and imbalanced turbulence, consistent with our results, but the $\bm{z^+}$ spectrum is steeper than $\bm{z^-}$. This inconsistency may be a result of the difference in the simulation setup, as \citet{perez2009role} and \citet{perez2012energy} implement driving forces for the turbulence while our simulations contain decaying turbulence.
In a recent MHD simulation of decaying strong turbulence \citep{yang2023energy}, the spectral slopes of $\bm{z^+}$ and $\bm{z^-}$ are roughly $-5/3$ and $-1$ respectively.
In addition, Parker Solar Probe observation shows that the $\bm{z^+}$ spectrum is mostly steeper than the $\bm{z^-}$ spectrum \citep{shi2021alfvenic}.
The discrepancy between our simulation result and these previous studies is still unclear. Nonetheless, we note that \citet{grappin2022modeling}, through a comprehensive parametric study with EBM simulations, find that the spectral slopes of $\bm{z^\pm}$ can be affected by various factors, including the initial spectral slopes and the turbulence strength.

\subsection{Higher-order statistics of the magnetic field}

\begin{figure*}[htb!]
    \centering
    \includegraphics[width=\hsize]{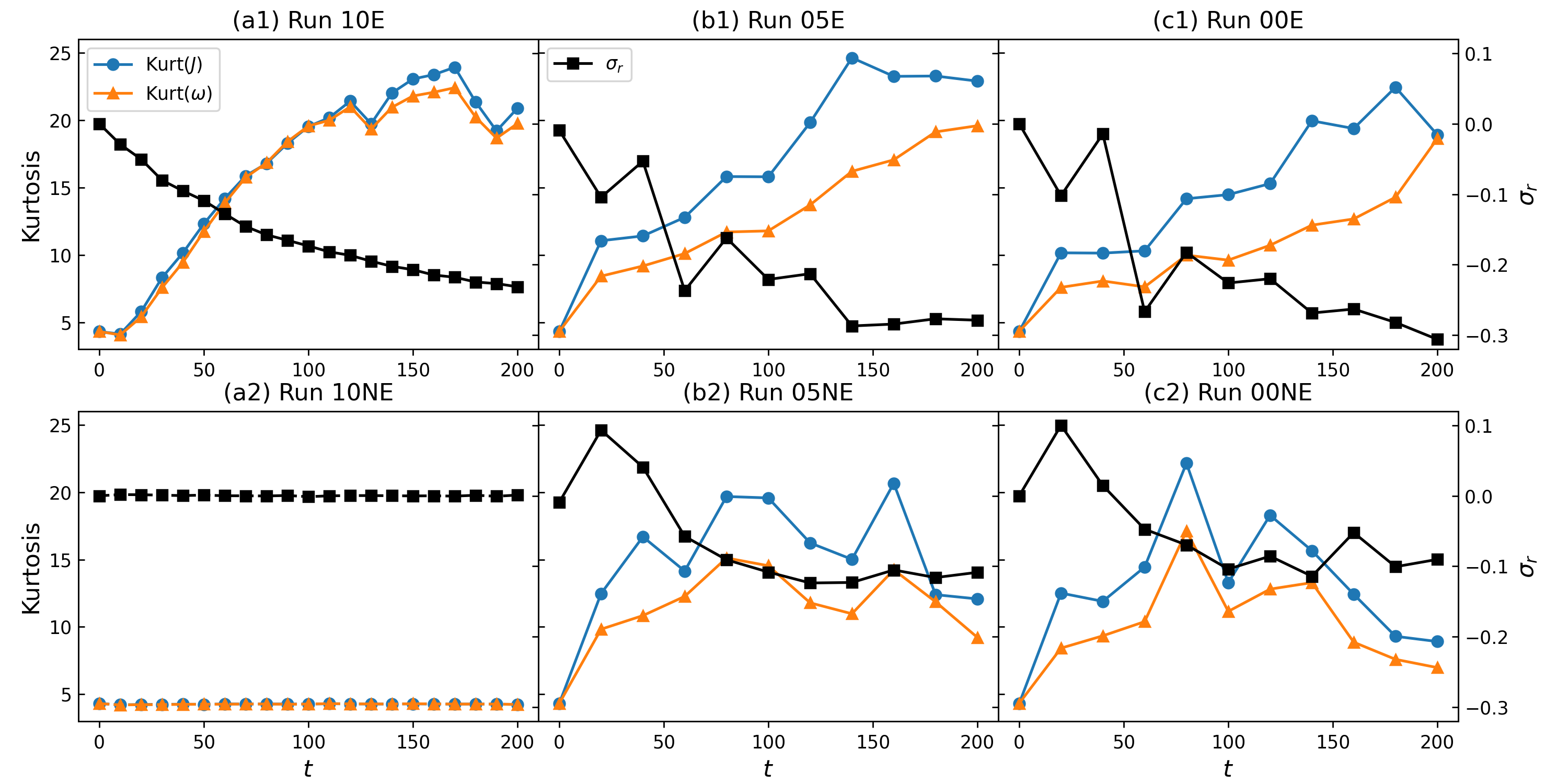}
    \caption{Time evolution of Kurtosis$(J)$ (blue circle), Kurtosis$(\omega)$ (orange triangle), and $\sigma_r$ (black square) in different runs. }
    \label{fig:time_evolution_Kurt_sigma_r}
\end{figure*}

It is well known that intermittency develops in MHD turbulence, generating local structures in magnetic field and velocity. A useful measure of the intermittency is the scaling exponents of structure functions. 
For homogeneous turbulence without intermittency, distribution of the fluctuations is typically assumed to be self-similar across different spatial scales, in which case the slope (``scaling exponent'') of the $q$-th order structure function is a linear function of $q$, i.e. ``mono-fractal.'' With intermittency, the slope becomes a nonlinear function of $q$, i.e. ``multi-fractal,'' due to the change of the distribution of the fluctuations as we move toward small scales. Observations have revealed that the magnetic field fluctuations in the solar wind are typically multi-fractal \citep[e.g.][]{sorriso1999intermittency,sioulas2022magnetic,palacios2022statistics}. 

In Figure \ref{fig:Sq_mag_Run10E}, we show the magnetic field structure functions $S_q(\bm{b},\bm{l})$ at the end of Run 10E as an example. 
For each curve, we apply linear fitting to the range $L/16 \leq l  \leq L/2$, which is marked by the yellow shades, and get the scaling exponents. The linear fitting result is shown by the black dashed lines.
In Figure \ref{fig:scaling_exponent}, we show the fitted scaling exponents at the end of different runs. As the Runs 05IC \& 00IC show similar results with Runs 05NE \& 00NE, they are not shown in this figure.
Here the blue curves with squares correspond to $l_x$, the orange curves with crosses correspond to $l_y$, and the green curves with triangles correspond to $l_z$. 
For references, the black dashed line shows $q/3$, which is the Kolmogorov turbulence model, the black dotted line shows $q/4$, which is the Iroshnikov-Kraichnan turbulence model, and the black dashed-dotted line is $1-q^{0.69}$, which is a multi-fractal intermittency model based on strong, balanced turbulence assumption \citep{chandran2015intermittency}. We note that the result for Run 10NE is unreliable because the structure functions barely develop a power-law in this run. 
Anisotropy among $x$, $y$, and $z$ axes as well as multi-fractality are clearly observed in all the runs. 
Inspecting the results on $l_z$, we find that the scaling exponents in Run 00NE roughly follow the prediction by \citep{chandran2015intermittency} as expected. Surprisingly, in Run 10E (strongly imbalanced turbulence with expansion), the scaling exponents also follow the prediction by \citep{chandran2015intermittency} which, however, is based on balanced turbulence assumption. 
Why Run 10E shows stronger multi-fractality than Runs 05E \& 00E is still unclear. It implies a complex interplay between the effect of nonlinear interaction and the effect of expansion on the evolution of intermittency.

Besides the scaling exponents, another useful quantification of intermittency is the Kurtosis of current density $J=|\nabla \times \bm{B}|$ and vorticity $\omega=|\nabla \times \bm{u}|$, which measure the strength of the intermittent current sheets and vortices. 
The Kurtosis of a variable, which quantifies the deviation of its probability distribution function from the Gaussian distribution, is defined as
\begin{displaymath}
    \mathrm{Kurtosis}(f) = \frac{\left< 
    f^4 \right>_{\bm{x}}}{\left< f^2 \right>_{\bm{x}}^2}
\end{displaymath}
where again $\left< \cdot \right>_{\bm{x}}$ stands for average over the simulation domain. We calculate the two quantities at the end of each run and write them in Figure \ref{fig:scaling_exponent}.
In Figure \ref{fig:time_evolution_Kurt_sigma_r}, we show the time evolution of Kurtosis$(J)$ (blue circle) and Kurtosis$(\omega)$ (orange triangle) in different runs. 
Again, Run 10NE does not show observable evolution due to the lack of nonlinear interaction. Runs 05IC \& 00IC are not shown because they show quite similar results with Runs 05NE \& 00NE.
Compare Kurtosis in different runs, we find that intermittency is obviously stronger in runs with expansion, possibly because of the selective decay of different components of the magnetic field and velocity which gives rise to small-scale structures \citep{dong2014evolution}. 
The Kurtosis in Run 10E grow much faster than Runs 05E \& 00E, and the growth of the Kurtosis in Run 05E is slightly faster than in Run 00E. This clearly shows that the evolution of intermittency is affected by $\sigma_c$.
Although the discrepancy between the evolution of Kurtosis in Runs 05NE and 00NE is not very pronounced, we expect to observe a slower evolution if the imbalance ($\sigma_c$) continues to increase, as implied by the stationary Kurtosis in Run 10NE (Panel (a2) of Figure \ref{fig:time_evolution_Kurt_sigma_r}).
Hence, the dependence of the Kurtosis growth rate on $\sigma_c$ is different in the expansion runs and the non-expansion runs. This is reminiscent of the result shown in Figure \ref{fig:scaling_exponent}, that is, the multifractality is stronger in Run 10E than in Runs 05E \& 00E, while it is stronger in Run 00NE than in Run 05NE.
This phenomenon is not fully understood yet and may imply a complex competition between the expansion effect and the nonlinear interaction in generating/dissipating the intermittency.
A theory of intermittency for imbalanced turbulence in the expanding solar wind and an observational study to compare the intermittency strength in solar wind streams with different $\sigma_c$ will be necessary.

\begin{figure*}[htb!]
    \includegraphics[width=\hsize]{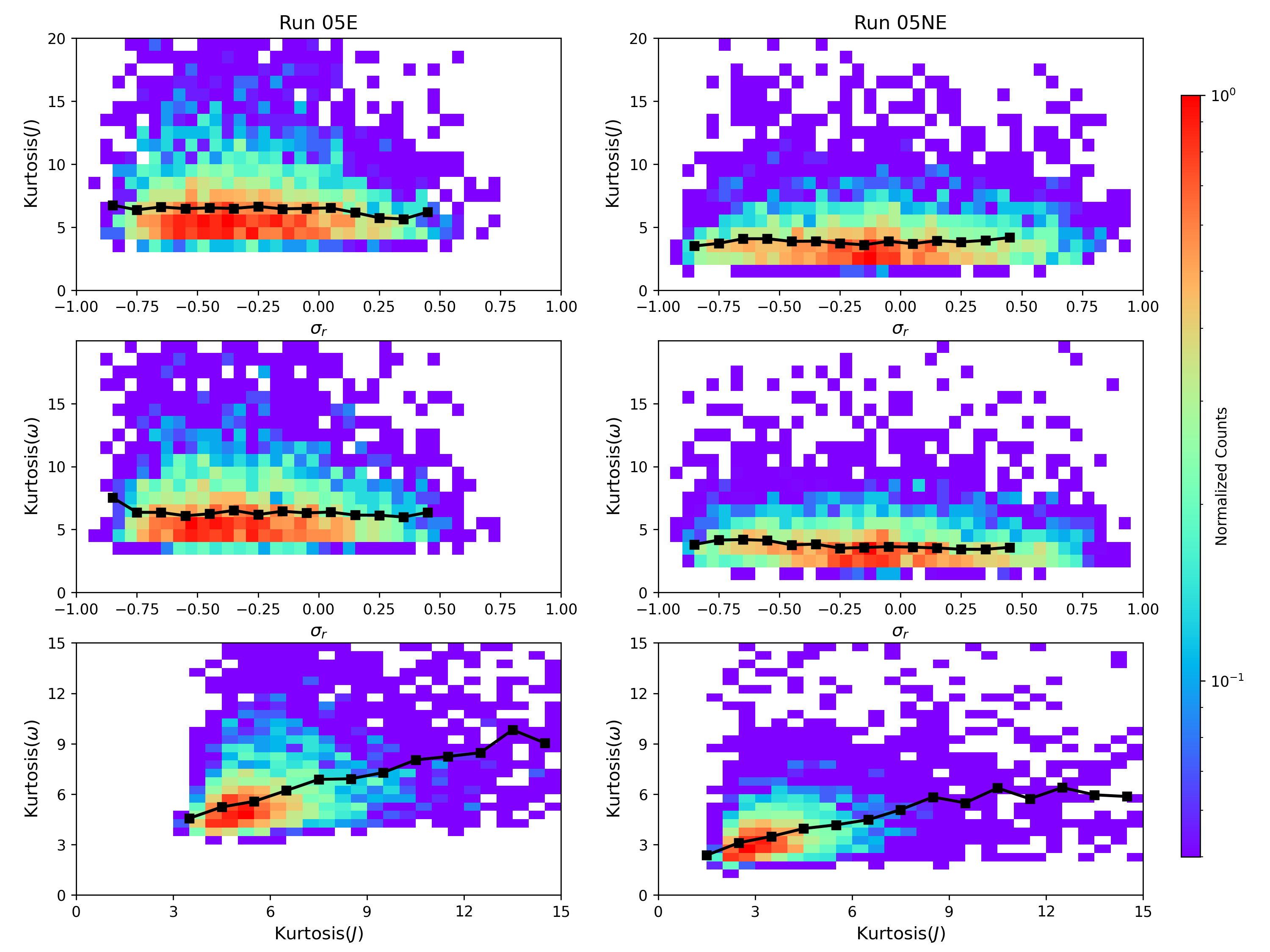}
    \caption{Probability distribution of Kurtosis$(J)$-$\sigma_r$ (top), Kurtosis$(\omega)$-$\sigma_r$ (middle), and Kurtosis$(\omega)$-Kurtosis$(J)$ (bottom) based on the last snapshot ($t=200$) of Run 05E (left) and Run 05NE (right). The simulation domain is divided into $16^3$ equal-size cubes and these quantities are calculated for each individual cube. Black lines are the median values of the $y$-axis values against the binned $x$-axis values. For Run 05E, the Pearson correlation coefficients are: C.C.(Kurt$(J)$-$\sigma_r$)=-0.03, C.C.(Kurt$(\omega)$-$\sigma_r$)=-0.02, and C.C.(Kurt$(\omega)$-Kurt$(J)$)=0.51. For Run 05NE, the Pearson correlation coefficients are: C.C.(Kurt$(J)$-$\sigma_r$)=0.02, C.C.(Kurt$(\omega)$-$\sigma_r$)=-0.06, and C.C.(Kurt$(\omega)$-Kurt$(J)$)=0.35.}
    \label{fig:kurtosis_residual}
\end{figure*}




\section{Discussion: Is residual energy related with intermittency?}\label{sec:discussion}
From Figure \ref{fig:time_evolution_Kurt_sigma_r}, one immediately notices that Kurtosis$(\omega)$ is smaller than Kurtosis$(J)$ in all the runs, implying magnetic structures are stronger than velocity structures, consistent with in-situ measurements by WIND \citep{bowen2018impact} and previous reduced-MHD simulations \citep{zhdankin2016scalings}. 
Consequently, one may conjecture that the negative residual energy is related to the intermittency. 
\citet{bowen2018impact} find that there is a negative correlation between Kurtosis$(J)$ and $\sigma_r$, which implies that the negative residual energy is likely related to the intermittent magnetic structures. 
In Figure \ref{fig:time_evolution_Kurt_sigma_r}, we show time evolution of $\sigma_r$ in black squares.
One can see that as the simulation goes, $\sigma_r$ decreases while the Kurtosis of both $J$ and $\omega$ increase.

Figure \ref{fig:time_evolution_Kurt_sigma_r} suggests that intermittent structures and negative residual energy are simultaneously generated as the turbulence evolves. 
However, whether the negative residual energy is produced by these intermittent structures is still unclear. 
To verify this point, we divide the simulation domain at the last frame ($t=200$) of each run evenly into $16 \times 16 \times 16$ cubes, i.e. each cube contains $32 \times 32 \times 32$ grid points. 
For each cube, we calculate Kurtosis($J$), Kurtosis($\omega$), and $\sigma_r$. Left and right columns of Figure \ref{fig:kurtosis_residual} show the probability distribution of the data points for Run 05E and Run 05NE respectively. Black curves are the median values of the $y$-axis values against the binned $x$-axis values. The other runs show similar results and hence are not shown here.
We calculate the Pearson correlation coefficient between each pair of parameters and these correlation coefficients are written in the Figure caption.
The bottom row of Figure \ref{fig:kurtosis_residual} shows that there is a positive correlation between Kurtosis($J$) and Kurtosis($\omega$), though the correlation coefficients (0.51 for Run 05E and 0.35 for Run 05NE) are not high.
This indicates that the intermittent structures in magnetic field and velocity are often co-located, but not always.
As shown by the top and middle rows, in both the two runs, Kurtosis($J$) and Kurtosis($\omega$) do not show significant correlation with $\sigma_r$, with nearly-zero correlation coefficients. Although the data points spread widely in $\sigma_r$, the median values of the Kurtosis are quite constant.
That is to say, at intermediate scales ($L/16$), regions with negative residual energy do not necessarily correspond to intermittent structures.
We have carried out the same analysis with the simulation domain divided into $8\times 8 \times 8$ cubes, i.e. for a larger spatial scale ($\sim L/8$), and the result (not shown here) is almost identical.

Thus, we conclude that, in our simulations, negative residual energy and intermittency are simultaneously generated as a result of turbulence evolution, but the causal relation between them seems to be weak. 
The reason is that the residual energy is concentrated at large scales (Figure \ref{fig:S2_ub}) while the intermittent structures are generated at smaller scales. 
Our result indicates that the negative residual energy is more likely produced by the wave-wave interaction \citep{boldyrev2012proceedingresidual,howes2013alfven} combined with the expansion effect. 
Intermittent structures may slightly contribute to the negative residual energy, considering the discrepancy between Kurtosis($\omega$) and Kurtosis($J$) in all the runs, but they are likely insignificant. 
Here we note that the weaker velocity intermittency than the magnetic intermittency is potentially due to the asymmetry between the momentum equation and the induction equation \citep{zhdankin2016scalings}. 
One can easily show that, in a 2D incompressible MHD system, the nonlinear term $ \bm{z^\mp} \cdot \nabla \bm{z^\pm}$ tends to contribute oppositely to the curl of the two Els\"asser variables $\bm{\omega^\pm} = \nabla \times \bm{z^\pm}$. 
Since $\bm{J} = \frac{1}{2}\left( \bm{\omega^+} -  \bm{\omega^-}\right)$ and $\bm{\omega} = \frac{1}{2}\left( \bm{\omega^+} +  \bm{\omega^-}\right)$, this asymmetry may lead to stronger current density than vorticity. However, a rigorous analysis of this problem still lacks and will be left for future work.

One possible explanation of the paradox between our simulations and satellite observations by \citep{bowen2018impact} is that \citet{bowen2018impact} adopted large time windows (one hour) to calculate these parameters and thus mixed large and small scales. 
Nonetheless, we note that due to artificial effects such as limited spatial resolution and lack of kinetic physics, the MHD simulations cannot capture all processes happening in the real solar wind. 
Moreover, besides the statistical analysis presented here, it would be beneficial to conduct a careful case study of the intermittent structures and their influences on the turbulence properties in MHD simulations in the future.



\section{Summary}\label{sec:summary}
We conducted a set of 3D MHD simulations of solar wind turbulence with intermediate strength ($|\delta \bm{b}| / B \sim 0.14$). The initialized fluctuations consist of counter-propagating Alfv\'en waves and has zero residual energy and varying normalized cross helicity. The key results are summarized below:
\begin{enumerate}
    \item Negative residual energy is always produced when nonlinear interaction takes effect, regardless of normalized cross helicity. The spherical expansion effect facilitates the generation of negative residual energy. 
    \item The magnetic field and velocity spectra are anisotropic and evolve differently. 
    The magnetic field spectrum has a quite universal perpendicular slope of $-3/2$ while the velocity spectrum is shallower. The negative residual energy is observed primarily in the perpendicular direction and has a spectrum $-E_r \propto k_\perp^{-2}$ in most runs.
    \item Spectral slope (along perpendicular direction) of $\bm{z^+}$ (outward) is quite universal and slightly shallower than $-3/2$, while the spectral slope of $\bm{z^-}$ (inward) highly depends on $\sigma_c$ when expansion effect is turned on such that the imbalanced turbulence has a steeper $\bm{z^-}$ spectrum. Without expansion, $\bm{z^-}$ spectrum has a slope of $-3/2$ for both the balanced runs (Runs 00NE \& 00IC) and imbalanced runs (Runs 05NE \& 05IC).
    \item Runs with expansion effect generate stronger intermittent structures in both magnetic field and velocity than the runs without expansion. The evolution of intermittency depends on $\sigma_c$ but the correlation between the intermittency and $\sigma_c$ is different in runs with expansion and in runs without expansion.
    \item Growth of negative residual energy is accompanied by the generation of intermittent structures. However, the causal relation between the negative residual energy and intermittency seems to be weak.
\end{enumerate}

We emphasize that the strength of turbulence in our simulations is smaller than what is observed in the young solar wind, where $|\delta \bm{B}|/B$ often reaches unity and thus magnetic ``switchbacks'' may form \citep{kasper2019alfvenic,bale2019highly,tenerani2020magnetic,tenerani2021evolution}. 
In addition, in the solar wind, the fluctuations are typically spherically polarized with $|B| = Const$ \citep{matteini20181,matteini2024alfvenic}. 
In numerical simulations, although there have been efforts to construct spherically polarized magnetic field in 3D \citep{valentini2019building,squire2022construction,johnston2022properties,shi2024analytic}, it is nontrivial to impose an constant-$|B|$ magnetic field with a specified spectral slope \citep{roberts2012construction}.



\vspace{1cm}
Acknowledgments: This study is supported by NSF SHINE \#2229566, NASA HTMS \#80NSSC20K1275, and NASA ECIP \#80NSSC23K1064. The numerical simulations are conducted on Extreme Science and Engineering Discovery Environment (XSEDE) EXPANSE at San Diego Supercomputer Center (SDSC) through allocation No. TG-AST200031, which is supported by National Science Foundation grant number ACI-1548562 \citep{Townsetal2014}. 

%

\vspace{5mm}


\software{Matplotlib \citep{Hunter2007Matplotlib}, NumPy \citep{harris2020array}, \texttt{LAPS} \citep{shi2024laps}}






\bibliography{references}{}

\begin{thebibliography}{}
\expandafter\ifx\csname natexlab\endcsname\relax\def\natexlab#1{#1}\fi
\providecommand{\url}[1]{\href{#1}{#1}}
\providecommand{\dodoi}[1]{doi:~\href{http://doi.org/#1}{\nolinkurl{#1}}}
\providecommand{\doeprint}[1]{\href{http://ascl.net/#1}{\nolinkurl{http://ascl.net/#1}}}
\providecommand{\doarXiv}[1]{\href{https://arxiv.org/abs/#1}{\nolinkurl{https://arxiv.org/abs/#1}}}

\bibitem[{Artemyev {et~al.}(2022)Artemyev, Shi, Lin, Nishimura, Gonzalez, Verniero, Wang, Velli, Tenerani, \& Sioulas}]{artemyev2022ion}
Artemyev, A., Shi, C., Lin, Y., {et~al.} 2022, The Astrophysical Journal, 939, 85

\bibitem[{Bale {et~al.}(2019)Bale, Badman, Bonnell, Bowen, Burgess, Case, Cattell, Chandran, Chaston, Chen, {et~al.}}]{bale2019highly}
Bale, S., Badman, S., Bonnell, J., {et~al.} 2019, Nature, 576, 237

\bibitem[{Belcher(1971)}]{belcher1971alfvenic}
Belcher, J. 1971, Astrophysical Journal, vol. 168, p. 509, 168, 509

\bibitem[{Belcher \& Davis~Jr(1971)}]{belcher1971large}
Belcher, J., \& Davis~Jr, L. 1971, Journal of Geophysical Research, 76, 3534

\bibitem[{Beresnyak \& Lazarian(2009)}]{beresnyak2009structure}
Beresnyak, A., \& Lazarian, A. 2009, The Astrophysical Journal, 702, 460

\bibitem[{Beresnyak \& Lazarian(2010)}]{beresnyak2010scaling}
---. 2010, The Astrophysical Journal Letters, 722, L110

\bibitem[{Boldyrev(2005)}]{boldyrev2005spectrum}
Boldyrev, S. 2005, The Astrophysical Journal, 626, L37

\bibitem[{Boldyrev {et~al.}(2012{\natexlab{a}})Boldyrev, Perez, \& Wang}]{boldyrev2012residual}
Boldyrev, S., Perez, J.~C., \& Wang, Y. 2012{\natexlab{a}}, arXiv preprint arXiv:1202.3453

\bibitem[{Boldyrev {et~al.}(2012{\natexlab{b}})Boldyrev, Perez, \& Zhdankin}]{boldyrev2012proceedingresidual}
Boldyrev, S., Perez, J.~C., \& Zhdankin, V. 2012{\natexlab{b}}, in AIP Conference Proceedings, Vol. 1436, American Institute of Physics, 18--23

\bibitem[{Bowen {et~al.}(2018)Bowen, Mallet, Bonnell, \& Bale}]{bowen2018impact}
Bowen, T.~A., Mallet, A., Bonnell, J.~W., \& Bale, S.~D. 2018, The Astrophysical Journal, 865, 45

\bibitem[{Bruno \& Carbone(2013)}]{bruno2013solar}
Bruno, R., \& Carbone, V. 2013, Living Reviews in Solar Physics, 10, 1

\bibitem[{Chandran(2018)}]{chandran2018parametric}
Chandran, B.~D. 2018, Journal of plasma physics, 84, 905840106

\bibitem[{Chandran {et~al.}(2015)Chandran, Schekochihin, \& Mallet}]{chandran2015intermittency}
Chandran, B.~D., Schekochihin, A.~A., \& Mallet, A. 2015, The Astrophysical Journal, 807, 39

\bibitem[{Chen {et~al.}(2013)Chen, Bale, Salem, \& Maruca}]{chen2013residual}
Chen, C., Bale, S., Salem, C., \& Maruca, B. 2013, The Astrophysical Journal, 770, 125

\bibitem[{Chen {et~al.}(2020)Chen, Bale, Bonnell, Borovikov, Bowen, Burgess, Case, Chandran, de~Wit, Goetz, {et~al.}}]{chen2020evolution}
Chen, C., Bale, S., Bonnell, J., {et~al.} 2020, The Astrophysical Journal Supplement Series, 246, 53

\bibitem[{Cranmer {et~al.}(2015)Cranmer, Asgari-Targhi, Miralles, Raymond, Strachan, Tian, \& Woolsey}]{cranmer2015role}
Cranmer, S.~R., Asgari-Targhi, M., Miralles, M.~P., {et~al.} 2015, Philosophical Transactions of the Royal Society A: Mathematical, Physical and Engineering Sciences, 373, 20140148

\bibitem[{Cranmer {et~al.}(2007)Cranmer, Van~Ballegooijen, \& Edgar}]{cranmer2007self}
Cranmer, S.~R., Van~Ballegooijen, A.~A., \& Edgar, R.~J. 2007, The Astrophysical Journal Supplement Series, 171, 520

\bibitem[{Dobrowolny {et~al.}(1980{\natexlab{a}})Dobrowolny, Mangeney, \& Veltri}]{dobrowolny1980properties}
Dobrowolny, M., Mangeney, A., \& Veltri, P. 1980{\natexlab{a}}, Solar and Interplanetary Dynamics, 143

\bibitem[{Dobrowolny {et~al.}(1980{\natexlab{b}})Dobrowolny, Mangeney, \& Veltri}]{dobrowolny1980fully}
---. 1980{\natexlab{b}}, Physical Review Letters, 45, 144

\bibitem[{Dong {et~al.}(2014)Dong, Verdini, \& Grappin}]{dong2014evolution}
Dong, Y., Verdini, A., \& Grappin, R. 2014, The Astrophysical Journal, 793, 118

\bibitem[{Dorfman {et~al.}(2024)Dorfman, Abler, Boldyrev, Chen, \& Greess}]{dorfman2024residual}
Dorfman, S., Abler, M., Boldyrev, S., Chen, C., \& Greess, S. 2024, arXiv preprint arXiv:2409.20442

\bibitem[{D’Amicis \& Bruno(2015)}]{d2015origin}
D’Amicis, R., \& Bruno, R. 2015, The Astrophysical Journal, 805, 84

\bibitem[{D’amicis {et~al.}(2019)D’amicis, Matteini, \& Bruno}]{d2019slow}
D’amicis, R., Matteini, L., \& Bruno, R. 2019, Monthly Notices of the Royal Astronomical Society, 483, 4665

\bibitem[{Gogoberidze {et~al.}(2012)Gogoberidze, Chapman, \& Hnat}]{gogoberidze2012generation}
Gogoberidze, G., Chapman, S.~C., \& Hnat, B. 2012, Physics of Plasmas, 19

\bibitem[{Goldreich \& Sridhar(1995)}]{goldreich1995toward}
Goldreich, P., \& Sridhar, S. 1995, The Astrophysical Journal, 438, 763

\bibitem[{Goldreich \& Sridhar(1997)}]{goldreich1997magnetohydrodynamic}
---. 1997, The Astrophysical Journal, 485, 680

\bibitem[{Grappin {et~al.}(1983)Grappin, Leorat, \& Pouquet}]{grappin1983dependence}
Grappin, R., Leorat, J., \& Pouquet, A. 1983, Astronomy and Astrophysics, 126, 51

\bibitem[{Grappin {et~al.}(2016)Grappin, M{\"u}ller, \& Verdini}]{grappin2016alfven}
Grappin, R., M{\"u}ller, W.-C., \& Verdini, A. 2016, Astronomy \& Astrophysics, 589, A131

\bibitem[{Grappin \& Velli(1996)}]{grappin1996waves}
Grappin, R., \& Velli, M. 1996, Journal of Geophysical Research: Space Physics, 101, 425

\bibitem[{Grappin {et~al.}(2022)Grappin, Verdini, \& M{\"u}ller}]{grappin2022modeling}
Grappin, R., Verdini, A., \& M{\"u}ller, W.-C. 2022, The Astrophysical Journal, 933, 246

\bibitem[{Halekas {et~al.}(2023)Halekas, Bale, Berthomier, Chandran, Drake, Kasper, Klein, Larson, Livi, Pulupa, {et~al.}}]{halekas2023quantifying}
Halekas, J., Bale, S., Berthomier, M., {et~al.} 2023, The Astrophysical Journal, 952, 26

\bibitem[{Harris {et~al.}(2020)Harris, Millman, van~der Walt, Gommers, Virtanen, Cournapeau, Wieser, Taylor, Berg, Smith, Kern, Picus, Hoyer, van Kerkwijk, Brett, Haldane, del R{\'{i}}o, Wiebe, Peterson, G{\'{e}}rard-Marchant, Sheppard, Reddy, Weckesser, Abbasi, Gohlke, \& Oliphant}]{harris2020array}
Harris, C.~R., Millman, K.~J., van~der Walt, S.~J., {et~al.} 2020, Nature, 585, 357, \dodoi{10.1038/s41586-020-2649-2}

\bibitem[{Heinemann \& Olbert(1980)}]{heinemann1980non}
Heinemann, M., \& Olbert, S. 1980, Journal of Geophysical Research: Space Physics, 85, 1311

\bibitem[{Hollweg(1974)}]{hollweg1974transverse}
Hollweg, J.~V. 1974, Journal of Geophysical Research, 79, 1539

\bibitem[{Howes \& Nielson(2013)}]{howes2013alfven}
Howes, G.~G., \& Nielson, K.~D. 2013, Physics of Plasmas, 20

\bibitem[{Huang {et~al.}(2022)Huang, Shi, Sioulas, \& Velli}]{huang2022conservation}
Huang, Z., Shi, C., Sioulas, N., \& Velli, M. 2022, The Astrophysical Journal, 935, 60

\bibitem[{Huang {et~al.}(2023)Huang, Sioulas, Shi, Velli, Bowen, Davis, Chandran, Matteini, Kang, Shi, {et~al.}}]{huang2023new}
Huang, Z., Sioulas, N., Shi, C., {et~al.} 2023, The Astrophysical Journal Letters, 950, L8

\bibitem[{Hunter(2007)}]{Hunter2007Matplotlib}
Hunter, J.~D. 2007, Computing in Science \& Engineering, 9, 90, \dodoi{10.1109/MCSE.2007.55}

\bibitem[{Iroshnikov(1964)}]{iroshnikov1964turbulence}
Iroshnikov, P. 1964, Soviet Astronomy, 7, 566

\bibitem[{Johnston {et~al.}(2022)Johnston, Squire, Mallet, \& Meyrand}]{johnston2022properties}
Johnston, Z., Squire, J., Mallet, A., \& Meyrand, R. 2022, Physics of Plasmas, 29

\bibitem[{Kasper {et~al.}(2019)Kasper, Bale, Belcher, Berthomier, Case, Chandran, Curtis, Gallagher, Gary, Golub, {et~al.}}]{kasper2019alfvenic}
Kasper, J.~C., Bale, S.~D., Belcher, J.~W., {et~al.} 2019, Nature, 576, 228

\bibitem[{Kraichnan(1965)}]{kraichnan1965inertial}
Kraichnan, R.~H. 1965, The Physics of Fluids, 8, 1385

\bibitem[{Lionello {et~al.}(2014)Lionello, Velli, Downs, Linker, Miki{\'c}, \& Verdini}]{lionello2014validating}
Lionello, R., Velli, M., Downs, C., {et~al.} 2014, The Astrophysical Journal, 784, 120

\bibitem[{Lithwick \& Goldreich(2003)}]{lithwick2003imbalanced}
Lithwick, Y., \& Goldreich, P. 2003, The Astrophysical Journal, 582, 1220

\bibitem[{Lithwick {et~al.}(2007)Lithwick, Goldreich, \& Sridhar}]{lithwick2007imbalanced}
Lithwick, Y., Goldreich, P., \& Sridhar, S. 2007, The Astrophysical Journal, 655, 269

\bibitem[{Magyar \& Nakariakov(2021)}]{magyar2021three}
Magyar, N., \& Nakariakov, V. 2021, The Astrophysical Journal, 907, 55

\bibitem[{Mallet \& Schekochihin(2017)}]{mallet2017statistical}
Mallet, A., \& Schekochihin, A.~A. 2017, Monthly Notices of the Royal Astronomical Society, 466, 3918

\bibitem[{Matteini {et~al.}(2018)Matteini, Stansby, Horbury, \& Chen}]{matteini20181}
Matteini, L., Stansby, D., Horbury, T., \& Chen, C.~H. 2018, The Astrophysical Journal Letters, 869, L32

\bibitem[{Matteini {et~al.}(2024)Matteini, Tenerani, Landi, Verdini, Velli, Hellinger, Franci, Horbury, Papini, \& Stawarz}]{matteini2024alfvenic}
Matteini, L., Tenerani, A., Landi, S., {et~al.} 2024, Physics of Plasmas, 31

\bibitem[{Meyrand {et~al.}(2023)Meyrand, Squire, Mallet, \& Chandran}]{meyrand2023reflection}
Meyrand, R., Squire, J., Mallet, A., \& Chandran, B.~D. 2023, arXiv preprint arXiv:2308.10389

\bibitem[{Montroll \& Shlesinger(1982)}]{montroll19821}
Montroll, E.~W., \& Shlesinger, M.~F. 1982, proceedings of the National Academy of Sciences, 79, 3380

\bibitem[{M{\"u}ller \& Grappin(2005)}]{muller2005spectral}
M{\"u}ller, W.-C., \& Grappin, R. 2005, Physical Review Letters, 95, 114502

\bibitem[{Palacios {et~al.}(2022)Palacios, Bourouaine, \& Perez}]{palacios2022statistics}
Palacios, J.~C., Bourouaine, S., \& Perez, J.~C. 2022, The Astrophysical Journal Letters, 940, L20

\bibitem[{Panasenco {et~al.}(2020)Panasenco, Velli, D’amicis, Shi, R{\'e}ville, Bale, Badman, Kasper, Korreck, Bonnell, {et~al.}}]{panasenco2020exploring}
Panasenco, O., Velli, M., D’amicis, R., {et~al.} 2020, The Astrophysical Journal Supplement Series, 246, 54

\bibitem[{Parashar {et~al.}(2020)Parashar, Goldstein, Maruca, Matthaeus, Ruffolo, Bandyopadhyay, Chhiber, Chasapis, Qudsi, Vech, {et~al.}}]{parashar2020measures}
Parashar, T., Goldstein, M., Maruca, B., {et~al.} 2020, The Astrophysical Journal Supplement Series, 246, 58

\bibitem[{Perez \& Boldyrev(2007)}]{perez2007weak}
Perez, J.~C., \& Boldyrev, S. 2007, The Astrophysical Journal, 672, L61

\bibitem[{Perez \& Boldyrev(2009)}]{perez2009role}
---. 2009, Physical review letters, 102, 025003

\bibitem[{Perez {et~al.}(2012)Perez, Mason, Boldyrev, \& Cattaneo}]{perez2012energy}
Perez, J.~C., Mason, J., Boldyrev, S., \& Cattaneo, F. 2012, Physical Review X, 2, 041005

\bibitem[{Phillips {et~al.}(2023)Phillips, Bandyopadhyay, McComas, \& Bale}]{phillips2023association}
Phillips, C., Bandyopadhyay, R., McComas, D.~J., \& Bale, S.~D. 2023, Monthly Notices of the Royal Astronomical Society: Letters, 519, L1

\bibitem[{R{\'e}ville {et~al.}(2020)R{\'e}ville, Velli, Panasenco, Tenerani, Shi, Badman, Bale, Kasper, Stevens, Korreck, {et~al.}}]{reville2020role}
R{\'e}ville, V., Velli, M., Panasenco, O., {et~al.} 2020, The Astrophysical Journal Supplement Series, 246, 24

\bibitem[{Rivera {et~al.}(2024)Rivera, Badman, Stevens, Verniero, Stawarz, Shi, Raines, Paulson, Owen, Niembro, {et~al.}}]{rivera2024situ}
Rivera, Y.~J., Badman, S.~T., Stevens, M.~L., {et~al.} 2024, Science, 385, 962

\bibitem[{Roberts(2012)}]{roberts2012construction}
Roberts, D.~A. 2012, Physical Review Letters, 109, 231102

\bibitem[{Shi {et~al.}(2024{\natexlab{a}})Shi, Tenerani, Rappazzo, \& Velli}]{shi2024laps}
Shi, C., Tenerani, A., Rappazzo, A.~F., \& Velli, M. 2024{\natexlab{a}}, Frontiers in Astronomy and Space Sciences, 11, 1412905

\bibitem[{Shi {et~al.}(2020)Shi, Velli, Tenerani, Rappazzo, \& R{\'e}ville}]{shi2020propagation}
Shi, C., Velli, M., Tenerani, A., Rappazzo, F., \& R{\'e}ville, V. 2020, The Astrophysical Journal, 888, 68

\bibitem[{Shi {et~al.}(2022)Shi, Velli, Tenerani, R{\'e}ville, \& Rappazzo}]{shi2022influence}
Shi, C., Velli, M., Tenerani, A., R{\'e}ville, V., \& Rappazzo, F. 2022, The Astrophysical Journal, 928, 93

\bibitem[{Shi {et~al.}(2024{\natexlab{b}})Shi, Velli, Toth, Zhang, Tenerani, Huang, Sioulas, \& van~der Holst}]{shi2024analytic}
Shi, C., Velli, M., Toth, G., {et~al.} 2024{\natexlab{b}}, The Astrophysical Journal Letters, 964, L28

\bibitem[{Shi {et~al.}(2021)Shi, Velli, Panasenco, Tenerani, R{\'e}ville, Bale, Kasper, Korreck, Bonnell, de~Wit, {et~al.}}]{shi2021alfvenic}
Shi, C., Velli, M., Panasenco, O., {et~al.} 2021, Astronomy \& Astrophysics, 650, A21

\bibitem[{Shoda {et~al.}(2019)Shoda, Suzuki, Asgari-Targhi, \& Yokoyama}]{shoda2019three}
Shoda, M., Suzuki, T.~K., Asgari-Targhi, M., \& Yokoyama, T. 2019, The Astrophysical Journal Letters, 880, L2

\bibitem[{Sioulas {et~al.}(2022{\natexlab{a}})Sioulas, Shi, Huang, \& Velli}]{sioulas2022preferential}
Sioulas, N., Shi, C., Huang, Z., \& Velli, M. 2022{\natexlab{a}}, The Astrophysical Journal Letters, 935, L29

\bibitem[{Sioulas {et~al.}(2022{\natexlab{b}})Sioulas, Velli, Chhiber, Vlahos, Matthaeus, Bandyopadhyay, Cuesta, Shi, Bowen, Qudsi, {et~al.}}]{sioulas2022statistical}
Sioulas, N., Velli, M., Chhiber, R., {et~al.} 2022{\natexlab{b}}, The Astrophysical Journal, 927, 140

\bibitem[{Sioulas {et~al.}(2022{\natexlab{c}})Sioulas, Huang, Velli, Chhiber, Cuesta, Shi, Matthaeus, Bandyopadhyay, Vlahos, Bowen, {et~al.}}]{sioulas2022magnetic}
Sioulas, N., Huang, Z., Velli, M., {et~al.} 2022{\natexlab{c}}, The Astrophysical Journal, 934, 143

\bibitem[{Sioulas {et~al.}(2023)Sioulas, Huang, Shi, Velli, Tenerani, Bowen, Bale, Huang, Vlahos, Woodham, {et~al.}}]{sioulas2023magnetic}
Sioulas, N., Huang, Z., Shi, C., {et~al.} 2023, The Astrophysical journal letters, 943, L8

\bibitem[{Sorriso-Valvo {et~al.}(1999)Sorriso-Valvo, Carbone, Veltri, Consolini, \& Bruno}]{sorriso1999intermittency}
Sorriso-Valvo, L., Carbone, V., Veltri, P., Consolini, G., \& Bruno, R. 1999, Geophysical Research Letters, 26, 1801

\bibitem[{Squire {et~al.}(2020)Squire, Chandran, \& Meyrand}]{squire2020situ}
Squire, J., Chandran, B.~D., \& Meyrand, R. 2020, The Astrophysical Journal Letters, 891, L2

\bibitem[{Squire \& Mallet(2022)}]{squire2022construction}
Squire, J., \& Mallet, A. 2022, Journal of Plasma Physics, 88, 175880503

\bibitem[{Tenerani {et~al.}(2021)Tenerani, Sioulas, Matteini, Panasenco, Shi, \& Velli}]{tenerani2021evolution}
Tenerani, A., Sioulas, N., Matteini, L., {et~al.} 2021, The Astrophysical Journal Letters, 919, L31

\bibitem[{Tenerani \& Velli(2017)}]{tenerani2017evolving}
Tenerani, A., \& Velli, M. 2017, The Astrophysical Journal, 843, 26

\bibitem[{Tenerani {et~al.}(2020)Tenerani, Velli, Matteini, R{\'e}ville, Shi, Bale, Kasper, Bonnell, Case, de~Wit, {et~al.}}]{tenerani2020magnetic}
Tenerani, A., Velli, M., Matteini, L., {et~al.} 2020, The Astrophysical Journal Supplement Series, 246, 32

\bibitem[{Towns {et~al.}(2014)Towns, Cockerill, Dahan, Foster, Gaither, Grimshaw, Hazlewood, Lathrop, Lifka, Peterson, {et~al.}}]{Townsetal2014}
Towns, J., Cockerill, T., Dahan, M., {et~al.} 2014, Computing in science \& engineering, 16, 62

\bibitem[{Valentini {et~al.}(2019)Valentini, Malara, Sorriso-Valvo, Bruno, \& Primavera}]{valentini2019building}
Valentini, F., Malara, F., Sorriso-Valvo, L., Bruno, R., \& Primavera, L. 2019, The Astrophysical Journal Letters, 881, L5

\bibitem[{Van~Ballegooijen \& Asgari-Targhi(2016)}]{van2016heating}
Van~Ballegooijen, A., \& Asgari-Targhi, M. 2016, The Astrophysical Journal, 821, 106

\bibitem[{Vasquez \& Hollweg(1998)}]{vasquez1998formation}
Vasquez, B.~J., \& Hollweg, J.~V. 1998, Journal of Geophysical Research: Space Physics, 103, 335

\bibitem[{Velli {et~al.}(1991)Velli, Grappin, \& Mangeney}]{velli1991waves}
Velli, M., Grappin, R., \& Mangeney, A. 1991, Geophysical \& Astrophysical Fluid Dynamics, 62, 101

\bibitem[{Verdini {et~al.}(2009)Verdini, Velli, Matthaeus, Oughton, \& Dmitruk}]{verdini2009turbulence}
Verdini, A., Velli, M., Matthaeus, W.~H., Oughton, S., \& Dmitruk, P. 2009, The Astrophysical Journal Letters, 708, L116

\bibitem[{Wang {et~al.}(2011)Wang, Boldyrev, \& Perez}]{wang2011residual}
Wang, Y., Boldyrev, S., \& Perez, J.~C. 2011, The Astrophysical Journal Letters, 740, L36

\bibitem[{Wu {et~al.}(2023)Wu, Huang, Wang, Yuan, He, \& Yang}]{wu2023intermittency}
Wu, H., Huang, S., Wang, X., {et~al.} 2023, The Astrophysical Journal Letters, 947, L22

\bibitem[{Yang {et~al.}(2023)Yang, He, Verscharen, Li, Bowen, Bale, Wu, Li, Wang, Zhang, {et~al.}}]{yang2023energy}
Yang, L., He, J., Verscharen, D., {et~al.} 2023, Nature Communications, 14, 7955

\bibitem[{Yokoi \& Hamba(2007)}]{yokoi2007application}
Yokoi, N., \& Hamba, F. 2007, Physics of Plasmas, 14

\bibitem[{Zhdankin {et~al.}(2016)Zhdankin, Boldyrev, \& Uzdensky}]{zhdankin2016scalings}
Zhdankin, V., Boldyrev, S., \& Uzdensky, D.~A. 2016, Physics of Plasmas, 23

\end{thebibliography}
\bibliographystyle{aasjournal}


\end{CJK*}
\end{document}